# Nonlinear magnetotransport shaped by Fermi surface topology and convexity in WTe$_2$

Pan He[1†], Chuang-Han Hsu[2,3†], Shuyuan Shi[1,2†], Kaiming Cai[1], Junyong Wang[2,3], Qisheng Wang[1], Goki Eda[2,3], Hsin Lin[4], Vitor M. Pereira[2,3] and Hyunsoo Yang[1,2]*

[1]*Department of Electrical and Computer Engineering, and NUSNNI, National University of Singapore, 117576, Singapore*

[2]*Centre for Advanced 2D Materials, National University of Singapore, 117546, Singapore*

[3]*Department of Physics, National University of Singapore, 117542 Singapore*

[4]*Institute of Physics, Academia Sinica, Taipei 11529, Taiwan*

[†]These authors contributed equally to this work. *e-mail: eleyang@nus.edu.sg

## Abstract

**The nature of Fermi surface defines the physical properties of conductors and many physical phenomena can be traced to its shape. Although the recent discovery of a current-dependent nonlinear magnetoresistance in spin-polarized non-magnetic materials has attracted considerable attention in spintronics, correlations between this phenomenon and the underlying fermiology remain unexplored. Here, we report the observation of nonlinear magnetoresistance at room temperature in a semimetal WTe$_2$, with an interesting temperature-driven inversion. Theoretical calculations reproduce the nonlinear transport measurements and allow us to attribute the inversion to temperature-induced changes in Fermi surface convexity. We also report a large anisotropy of nonlinear magnetoresistance in WTe$_2$, due to its low symmetry**

**of Fermi surfaces. The good agreement between experiments and theoretical modeling reveals the critical role of Fermi surface topology and convexity on the nonlinear magneto-response. These results lay a new path to explore ramifications of distinct fermiology for nonlinear transport in condensed-matter.**

## Introduction

Layered transition metal dichalcogenides are an emerging class of materials with novel physical phenomena and a wide range of potential applications[1, 2]. Among them, semimetal tungsten ditelluride ($WTe_2$) especially showed an extremely large and unsaturated magnetoresistance (MR)[3], which was attributed to a Fermi surface with perfectly compensated electron and hole pockets[3-5]. Up to now, the MR investigated in $WTe_2$ appears in the linear region and the exploration of a current-dependent nonlinear magnetoresistance is lacking. In this context, a nonlinear magnetoresistance (NLMR) that scales linearly with the applied electric and magnetic fields has been recently discovered independently in a polar semiconductor[6] and a topological insulator[7] with spin-momentum locked bands, and has led to a surge of interest within condensed matter physics towards understanding its underlying mechanisms[8-10]. In $WTe_2$, the large spin-orbit coupling (SOC) and broken inversion symmetry lift the spin degeneracy, as observed in spin- and angle-resolved photoemission spectroscopy (ARPES)[11-13]. Taking advantage of the large SOC and low crystalline symmetry, $WTe_2$ was recently demonstrated as an intriguing spin-source for generating out-of-plane anti-damping torques to an adjacent magnetic material[14]. To further explore ramifications of the spin-polarized bands in $WTe_2$, and their interplay with the richly structured Fermi surface[15-17], we investigate in detail the spin-dependent nonlinear magnetotransport.

WTe$_2$ exhibits properties remarkably sensitive to temperature as a combined result of nearly perfect carrier compensation at low temperatures with a hole-suppressing Lifshitz transition at ~160 K[18]. Accordingly, small temperature variations cause essential changes in the Fermi surface, as revealed by ARPES[5, 18]. In addition, WTe$_2$ undergoes a topological transition from a Weyl to a topologically trivial semimetal at ~70 K[17, 19]. While an understanding of the Fermi surface is critical to describe and explore many technologically promising physical phenomena, such as the oscillatory exchange coupling or superconductivity[20], and even though previous studies investigated the role of spin textures in both **k** and real spaces[6, 7, 9, 10], the fermiological implications and opportunities remain unexplored in the context of nonlinear magnetotransport. With such a rich and easily tunable Fermi surface, WTe$_2$ is an excellent platform to investigate these questions.

In this work, we report the observation of a current-dependent NLMR in WTe$_2$ that scales with the first power of both the applied electric current and magnetic field[6, 7]. This is in striking contrast with the conventional (linear) MR so far characterized in WTe$_2$, which is current-independent and quadratic in the magnetic field[3, 4]. Interestingly, the NLMR shows a temperature-driven inversion and a significant anisotropy along different crystallographic axes. Our experimental results are reproduced qualitatively by theoretical modeling that combines ab-initio band structure calculations with a semiclassical calculation of the magneto-response. The calculations reveal that the spin-polarized electronic structure evolves with the magnetic field, giving rise to the measured spin-dependent NLMR. Furthermore, we establish that the inversion of NLMR arises from a transition in the Fermi contours from convex to concave, whereas the giant anisotropy is due to the low symmetry of the Fermi surface. Therefore, we establish here a close relationship between Fermi surface topology, convexity and the nonlinear magnetotransport response. These results also demonstrate that fine tuning of the Fermi level is critical in controlling the nonlinear magneto-transport in semimetals.

## Results

**Sample and characterization**

At ambient conditions, bulk $WTe_2$ is in the orthorhombic $T_d$ phase[21] where, as a result of strong octahedral distortions and the displacement of the metal within each octahedron, the W ions organize as effective zigzag chains along the *a* axis and sandwiched between Te layers as shown in Fig. 1a. The band structure calculated from the density functional theory (DFT) is plotted in Fig. 1b, and agrees with previous calculations[18, 22]. In Fig. 1c, we show a schematic representation of the Fermi surface with only electron pockets at different carrier filling that illustrates how the sign of the NLMR depends on whether the Fermi surface consists of a concave set or a convex set. In bulk $WTe_2$, the Fermi surface at 0 K consists of multiple hole and electron pockets, which can be seen below.

Our $WTe_2$ flakes were obtained by mechanical exfoliation on $Si/SiO_2$ substrates and subsequently patterned into Hall devices for transport measurements (see Methods). Multiple devices were prepared with the current channels (that define our *x* direction) designed along different crystallographic directions of the underlying $WTe_2$. The typical temperature dependence of their resistivity ($\rho$) is shown in Fig. 1d, and is consistent with previous reports[3, 13]. To first order in the applied current *I*, the longitudinal resistance *R* can be expressed as $R = R_0 + R'I$, where $R_0$ is the current-independent (linear) resistance and $R'I$ is the current-dependent (nonlinear) resistance. In order to study the nonlinear magnetotransport, a low-frequency a.c. current $I_\omega = I\sin(\omega t)$ was applied through the devices (Fig. 1e, f), and we recorded the second harmonic longitudinal voltage $V_{2\omega} = \frac{1}{2}R'I^2\sin(2\omega t - \pi/2)$ using lock-in techniques. The second harmonic resistance $R_{2\omega}$ ($R_{2\omega} \equiv \frac{1}{2}R'I$), equivalent to half of the nonlinear resistance $R'I$, is used to quantify the nonlinear magnetotransport response (which we verified to be independent of the driving frequency)[23]. The measurements were performed while rotating the applied magnetic field **H** in the *xy* plane of the film (out-of-plane misalignment evaluated to be less than 1°) at an angle $\varphi$ with the current direction, as illustrated in Fig. 1f. We note that, unlike the colossal linear MR which is the strongest under perpendicular fields[3-5], the nonlinear magnetoresponse is maximized under in-plane fields due to the planar spin texture of bulk $WTe_2$, as we discuss below.

**Nonlinear magnetoresistance in WTe₂**

We observe a sinusoidal dependence of the second-harmonic resistance $R_{2\omega}$ on the angle $\varphi$ between the magnetic field and the current direction. Fig. 2a shows that the period is 360° with a maximum when the field is orthogonal to the current ($\varphi = 90°$ or 270°), and approaches zero when they are collinear ($\varphi = 0°$ or 180°). Hence, unlike the first harmonic resistance $R_\omega$ (see Supplementary Fig. 1), $R_{2\omega}$ changes sign upon reversal of the magnetic field. To further characterize the NLMR, $R_{2\omega}$ was measured at different *I* and *H* at 300 K, and the same dependence $R_{2\omega} = -\Delta R_{2\omega}\sin\varphi$ was observed. Fig. 2b shows that the extracted amplitude $\Delta R_{2\omega}$ increases linearly with $I$ and, at the same time, scales linearly with $H$, as plotted in Fig. 2c (the behavior shown in Figs. 2a-c persists down to 2 K). This contrasts with the field dependence of the linear resistance $R_\omega$ which grows quadratically with $H$ (Supplementary Fig. 1)[3]. These observations extend the class of systems bearing current-dependent nonlinear magnetotransport to the layered semimetal $WTe_2$, where it is observable at room temperature.

**Temperature-driven inversion of the nonlinear magnetoresistance**

Motivated by the strong temperature dependence of its other known transport properties[5, 17-19], we study in detail the NLMR of $WTe_2$. Measurements of the $\varphi$-dependent $R_{2\omega}$ at different temperatures (Supplementary Fig. 2) reveal that, upon lowering $T$ from room temperature, the initially positive amplitude $\Delta R_{2\omega}$ undergoes a gradual reduction until it reaches zero at $T \approx 140$ K, at which point it changes sign and progressively increases to large negative values as the temperature reduces. This behavior is shown in Fig. 2d, which is distinct from that of the linear MR (see Supplementary Fig. 1). We note that the inversion in $R_{2\omega}$ occurs at a temperature close to that of the reported Lifshitz transition in bulk $WTe_2$[18], where the Fermi surface topology changes. In order to evaluate changes of the electronic structure with temperature[18, 19], we measure the Hall resistance $R_{xy}$ in Fig. 2e which displays a gradual deviation from the linear field dependence at low temperatures. By fitting $R_{xy}$ ($H$) according to a two-carrier model[19], we extract the hole (electron) density, which increases (decreases) upon lowering the temperature (Fig. 2f). This is consistent with the strong sensitivity of the chemical potential to temperature changes in $WTe_2$[5, 18]. As we show below, the temperature-driven changes in the chemical potential are reflected not only in variations of the size of the electron and hole pockets[5, 18], as schematically shown in the inset of Fig. 2d, but also in the convexity of the Fermi contours. The latter effect turns out to be crucial to drive the inversion of the NLMR.

**Giant crystal anisotropy of nonlinear magnetoresistance**

The strong local distortion of the W ions in the $T_d$ structure causes them to arrange along one-dimensional zigzag chains parallel to the *a* axis within each monolayer, and this imparts a strong electronic anisotropy[3, 5]. To investigate its consequences in the NLMR $R_{2\omega}$, we pattern circular Hall devices as shown in Fig. 3a, in which currents can be applied along different crystallographic directions of the same $WTe_2$ device. One of the current channels is chosen parallel to the *a* axis of $WTe_2$ (Fig. 1a), as identified by polarized Raman spectra[24] (see Supplementary Fig. 3). The longitudinal (linear) resistivity $\rho$ is plotted in Fig. 3b and, as expected, is anisotropic with values along the *a* axis ~3 times lower than those along $b^{25}$. The behavior of the second harmonic $R_{2\omega}$ is, however, more interesting as plotted in Fig. 3c for the four different directions of current flow. The magnitude of $R_{2\omega}$ is much stronger along *b* compared to that of the *a* direction, with a *b*/*a* magnitude ratio ~18 at 2 K and ~5 at 300 K. Moreover, the sign change in $R_{2\omega}$ is also sensitive to the current direction in relation to the crystallographic axes: the sign inversion with temperature is clear when the current flows along the *b* axis, and inversion is absent when it flows parallel to *a*. We define $\chi = 2 R_{2\omega} / (R_\omega I H)$ which characterizes the NLMR under unit electric voltage and magnetic field shown in Fig. 4d, and note that $\chi$ reaches up to 0.04 and $-0.22$ $\Omega$ $V^{-1}$ $T^{-1}$ (or $A^{-1}$ $T^{-1}$) along the *a* and *b* directions at 2 K, respectively. Such a large anisotropy in the NLMR constitutes a record, and has not been observed in other materials[6, 7, 9, 10]. The discovery of this giant anisotropy from a nonlinear magnetotransport perspective enriches our understanding of the anisotropic Fermi surface in WTe$_2$ as we discuss below.

**Theoretical modeling**

To gain insight into the observed NLMR in WTe$_2$, we calculate the longitudinal second-order current density, $J_{xx}^{(2)}$, in the presence of an external magnetic field **H** perpendicular to **E** (Supplementary Note 1), and use the fact that $J_{xx}^{(2)} \propto -R_{2\omega}$ (see Supplementary Note 2). The calculation relies on a Wannier tight-binding Hamiltonian that reproduces all the details of the DFT band structure of bulk WTe$_2$ given in Fig. 1b, including the relative positions of hole and electron pockets, as well as the flat bands immediately below the Fermi energy that were discussed in Ref. [5]. It is important to note that, in semimetallic WTe$_2$, the Fermi level $\mu$ is known to change by ~50 meV between 40 and 160 K[18]. To corroborate this, we have calculated the hole ($n_h$) and electron ($n_e$) densities at different $\mu$ (Supplementary Note 3) which are shown with the experimental ones in Fig. 2f. We establish the relation between $T$ (bottom axis in Fig. 2f) and $\mu$ (top axis in Fig. 2f) by comparing the calculated $n_e$ and $n_h$ with the experimental data.

Fig. 4a displays the dependence of calculated $J^{(2)}$ on $\mu$ at different temperatures for current flowing along the $b$ direction. The sign of $J_{bb}^{(2)}$ changes from positive to negative with increasing $\mu$, and the inversion threshold is mostly insensitive to the thermal broadening. This indicates that the experimentally observed inversion in $R_{2\omega}$ is primarily due to changes in the Fermi level. For a direct comparison, we plot in Fig. 4d the calculated ratio $J_{ii}^{(2)}/[J_{ii}^{(1)}]^2 \propto -\chi$ (Supplementary Note 2) versus $T$ for the current along the $a$ and $b$ axes ($i = a, b$). The calculated result captures the overall qualitative behavior of the experimental data in Fig. 3d, with the sign inversion and large crystalline anisotropy. In addition, the theoretical $J^{(2)}$ in Fig. 4e reproduces the experimental sinusoidal dependence on $\varphi$ (Fig. 2a), and the linearity of $J^{(2)}$ in Fig. 4f with respect to the magnetic field (Fig. 2c). Another significant implication from the calculations is that the sign inversion of $J^{(2)}_{bb}$ is dominated by the electron carriers as can be seen from the carrier-resolved $J^{(2)}_{bb}$ in Supplementary Fig. 4. This indicates that the electron pockets are responsible for the sign change.

We now discuss the role of the Fermi surface topology to understand the non-trivial transport phenomena, NLMR in $WTe_2$. The calculated Fermi contours are plotted in Figs. 4b and 4c at two Fermi levels associated with opposite $J^{(2)}$ in Fig. 4a. In addition to the suppression of hole pockets, we find a distinctive change in the 3-dimensional (3D) convexity of the Fermi surface: the Fermi contours contain portions with a concave (CC) shape at $\mu = 0$ (Fig. 4b), which evolves into entirely convex (CV) contours at $\mu = 120$ meV (Fig. 4c). Since $J^{(2)}$ is governed by both the local band velocity and curvature, the Fermi surface convexity should determine the sign of the nonlinear current. To further illustrate this characteristic observed consistently both in the experiments and the calculations, we build a simplified quasi-bulk model Hamiltonian[26] based on symmetry (see Supplementary Fig. 7 and Supplementary Note 4). Reflecting what is seen in the DFT-derived band structure (Figs. 4b and 4c), this model yields electron pockets whose Fermi surface convexity changes with the chemical potential. We calculate $J^{(2)}_{bb}$ based on this simpler, stripped-down Hamiltonian and see the same sign inversion at a threshold $\mu$, as is clear from Fig. 4g. Moreover, the inversion threshold is directly correlated to the transition of the Fermi contour from CV to CC, as illustrated in Figs. 4h and 4i: Whereas at low $\mu$ (panel 4i) the longitudinal dispersion is parabolic, at higher $\mu$ (panel 4h) it becomes shaped like a Mexican hat, a direct manifestation of the CC of the Fermi contour. Thus, it can be concluded that a transition from CV to CC is a necessary condition for the observation of a sign inversion in $J^{(2)}$, as intuitively expected from the analytical expression for $J^{(2)}_{bb}$ which depends on the effective mass only along the $b$ direction (see Supplementary Note 5).

This establishes the convexity of the Fermi surface as the driving mechanism behind the sign inversion in $J^{(2)}$ with temperature and indicates that the NLMR is a simple transport observable to electrically monitor the variation of the Fermi level with temperature in this system. Overall, our results confirm the critical role of topology and convexity of the Fermi surface on the exotic nonlinear magnetotransport of $WTe_2$.

## Discussion

In semimetallic $WTe_2$, we have demonstrated that it exhibits a NLMR at room temperature, which is sensitive to temperature including a sign reversal, and strongly anisotropic. These properties critically depend on the Fermi surface morphology. In $WTe_2$, the strong temperature dependence arises from an unusually large thermal shift of the chemical potential.

Nonlinear magneto-currents associated with strong SOC and Fermi surfaces with nontrivial spin textures are a nascent field of research. Here we have shown the ability to control their magnitude, and especially their sign, with either magnetic fields, or temperature in $WTe_2$. In addition, doping[22], pressure[27], electrostatic gating[15, 28, 29] and film thickness[30-32] are known to be effective to tune the Fermi surface of WTe$_2$. Reducing the thickness was found to modify the electronic structure of WTe$_2$ films[32], and flakes with different thicknesses display different $R_{2\omega}$ inversion temperatures (Supplementary Fig. 9). The extreme case of monolayer 1T'-WTe$_2$ was recently shown to be a two-dimensional topological insulator with an insulating bulk and a topologically nontrivial metallic edge[26, 33-35], although the nonlinear magneto-transport at the monolayer edge is still an open question. On the other hand, bulk WTe$_2$ was predicted to be a type-II Weyl semimetal[36], but the nonlinear magneto-response when the Fermi energy of a Weyl semimetal is tuned to the Weyl nodes remains unknown.

All these factors suggest ample space for further tunability and possibly even richer nonlinear, spin-dependent features in the charge and spin transport of WTe$_2$. Moreover, changes in the topology and convexity of the Fermi surface are widely observed in other materials, such as the Dirac semimetal ZrTe$_5$[37] and LaAlO$_3$/SrTiO$_3$[38] at the Lifshitz transition. Details of the Fermi surface are also very sensitive to the strength of SOC in Weyl semimetals of the transition metal monopnictide family[39], such as TaAs[40], TaP[41] and NbP[39]. We anticipate the existence of a NLMR in these materials as well which, if confirmed, could establish this effect as a transport-only probe of topological and other types of Fermi surface transitions.

**Methods**

**Sample preparation.** WTe$_2$ thin film flakes with different thicknesses were cleaved from a bulk WTe$_2$ single crystal (HQ Graphene) by mechanical exfoliation onto a 300 nm Si/SiO$_2$ substrate with alignment markers. Subsequently, a capping layer of SiO$_2$ (4 nm) was deposited on them as a protection layer to eliminate material degradation during subsequent fabrication processes. Hall bar devices with a conventional rectangular (as well as radial multi-terminal) geometry were fabricated using the standard photolithography (E-beam lithography) method and Ar$^+$-milling. Prior to the Cr (30 nm)/Au (60 nm) electrode deposition using magnetron-sputtering, the contact regions were treated with Ar$^+$-etching to remove the SiO$_2$. Ohmic contacts were confirmed by *I-V* measurements. The crystallographic direction of the current channel was confirmed by performing polarized Raman measurements with a misalignment $< \pm 5°$. The flake thickness and surface morphology were measured by an atomic force microscope.

**Electrical measurements.** Hall bar devices were wire-bonded to a rotatable sample holder and installed in a physical property measurement system (PPMS, Quantum Design) for transport measurements in the temperature range 2–300 K. We performed the measurements of a.c. harmonic resistances using a Keithley 6221 current source and Stanford Research SR830 lock-in amplifiers. During the measurements, a constant amplitude sinusoidal current with a frequency of 21 Hz was applied to the devices, and the in-phase (0°) first harmonic $V_\omega$ and out-of-phase (-90°) second harmonic $V_{2\omega}$ longitudinal voltage signals were measured simultaneously by two lock-in amplifiers.

**Theoretical modeling.** The starting point for our calculations of the carrier density (Fig. 2f) and nonlinear current (Figs. 4a-f) is the electronic band structure of bulk $WTe_2$ obtained within DFT. From the ab-initio results, we obtain an accurate Wannier tight-binding Hamiltonian representation that describes the electronic structure of this system, and which is subsequently used in our transport-related calculations. We used the DFT-derived tight-binding that has been previously employed to predict Fermi arcs in this material in reference [22]. The observable transport properties are calculated from Boltzmann response theory using the tight-binding Hamiltonian (Supplementary Note 1). In order to isolate analytically the Fermi surface convexity as the underlying cause of the sign-change in the nonlinear MR, we also considered the effective $\mathbf{k} \cdot \mathbf{p}$ Hamiltonian dictated by symmetry for this system, as described in Supplementary Notes 4 and 5. The results shown in Figs. 4g-i have been obtained with this approximated Hamiltonian; all other calculations are based on the full DFT-derived tight-binding Hamiltonian.

**Data availability.** The data that support the plots within this paper and other findings of this study are available from the corresponding author upon reasonable request.

**Acknowledgments**

The work was partially supported by the National Research Foundation (NRF), Prime Minister's Office, Singapore, under its Competitive Research Programme (CRP award no. NRFCRP12-2013-01). Numerical calculations were performed at the HPC facilities of the NUS Centre for Advanced 2D Materials.

**Author contributions**

P.H., S.S. and H.Y. designed the experimental study. S.S. and P.H. fabricated devices, P.H. performed the transport measurements and analyzed the data. K.C., J.Y.W, Q.W. and G.E. helped in sample characterization. C.-H.H., V.M.P and H.L. designed the theoretical methodology with C.H.H performing all the numerical calculations. All authors discussed the results. P.H., C.-H.H., V.M.P, and H.Y. wrote the manuscript. H.Y. supervised the project.

**Additional information**

Supplementary information is available in the online version of the paper. Reprints and permissions information is available online at www.nature.com/reprints. Correspondence and requests for materials should be addressed to H.Y.

**Competing interests**

The authors declare no competing interests.

**Figure captions**

**Fig. 1 | Crystal & band structures and nonlinear magnetotransport in WTe$_2$. a**, Crystal structure of the layered WTe$_2$ and its crystalline directions **b**, The calculated band structure of bulk WTe$_2$, where the high symmetry **k** points are indicated in the 3D Brillouin zone sketched underneath. A magnified plot of the band dispersion is also shown to emphasize details of the electron and hole pockets. **c**, Schematic Fermi surfaces at two representative Fermi energies illustrating the distinct convexity (see Fig. 4b for the full realistic Fermi surfaces and contours of bulk WTe$_2$ at $T = 0$ K). Under a magnetic field **H** perpendicular to the applied electric field **E** (along $b$ axis), a nonlinear charge current density $J_{bb}^{(2)}$ at the second order in **E** is generated along the $b$ direction. $J_{bb}^{(2)}$ has opposite sign at the two Fermi energies. **d**, Temperature-dependent resistivity $\rho$ with a thickness of 34 nm WTe$_2$. **e**, A typical optical image of Hall bar devices. **f**, Schematic of harmonic MR measurements while rotating **H** in the $xy$ plane at an angle $\varphi$ with the current.

**Fig. 2 | Nonlinear magnetoresistance with a temperature-driven inversion in WTe$_2$. a**, Angular-dependent second harmonic resistance $R_{2\omega}$ measured at $T = 300$ K for a 34 nm thick WTe$_2$ device with the current applied at 45° from the $a$ axis of WTe$_2$ crystal. The solid line is a sinusoidal fit ($-\Delta R_{2\omega}\sin\varphi$) to the data. A vertical offset was subtracted for clarity. **b**, **c**, Current $I$ (**b**) and magnetic field $H$ (**c**) dependence of the sinusoidal amplitude $\Delta R_{2\omega}$ at room temperature. The solid lines are linear fits to the data. **d**, $\Delta R_{2\omega}$ extracted at different temperatures under $H = 14$ T and $I = 1$ mA. The insets illustrate the position of the Fermi level in different temperature ranges. 'e' and 'h' indicates the electron and hole pocket, respectively. **e**, Field dependence of the Hall resistance $R_{xy}$ at three representative temperatures (red lines are to guide the eye). **f**, Experimental temperature dependence of

the electron $n_e$ and hole $n_h$ densities (points, bottom and left axes), overlaid with the Fermi level dependence of $n_e$ and $n_h$ as calculated ab-initio (lines, top and right axes).

**Fig. 3 | Crystal anisotropy of nonlinear magnetoresistance. a**, Optical image of a circular Hall bar device with a flake thickness of 13.6 nm with arrows indicating the *a* and *b* directions. **b**, Temperature dependence of the channel resistivity $\rho$ along different crystallographic directions. **c,d**, Temperature dependence of the nonlinear magnetoresistance $R_{2\omega}$ (at $\varphi = 270°$) normalized under unit current and magnetic field (**c**) and normalized under unit electric voltage and magnetic field (**d**) along different crystallographic orientations.

**Fig. 4 | Theoretical nonlinear charge current and Fermi surface. a**, Calculated longitudinal second order current density $J_{bb}^{(2)}$ versus Fermi level $\mu$ for current flowing along the *b* axis with a Zeeman energy of 0.1 meV. **b, c**, Calculated Fermi surfaces of bulk WTe$_2$ at $\mu = 0$ and 120 meV, respectively, with three horizontal cuts shown underneath. 'e' and 'h' indicates the electron and hole pocket, respectively. Based on the result of Fig. 2f, these two chemical potentials correspond to $T = 0$ K and 300 K, respectively. **d**, Temperature dependence of the ratio $-J_{ii}^{(2)}/[J_{ii}^{(1)}]^2 \propto \chi$ for the current applied along the *a* and *b* axes. The *T* axis takes into account both the thermal broadening by the Fermi-Dirac distribution and the Fermi level shift. **e, f**, Variation of the calculated $J_{bb}^{(2)}$ at 300 K with magnetic field angle $\varphi$ (**e**) and with the field intensity at $\varphi = 90°$ (**f**). Dashed lines are, respectively, the function $\sin(\varphi)$ and a linear fit to the calculated points. The above results in **a-f** are calculated based on the Wannier Hamiltonian of bulk WTe$_2$, which reproduces the DFT band structure shown in Fig. 1b. **g,** The calculated second order current density

$J_{bb}^{(2)}$ versus the Fermi level $\mu$ for the simplified quasi-bulk tight-binding model. **h, i**, The energy dispersion and Fermi surfaces at $\mu$ = 200 and 75 meV, respectively, which correspond to the values marked by the vertical dashed lines in panel **g**. Calculated currents are presented in arbitrary units in all panels (Supplementary Note 1).

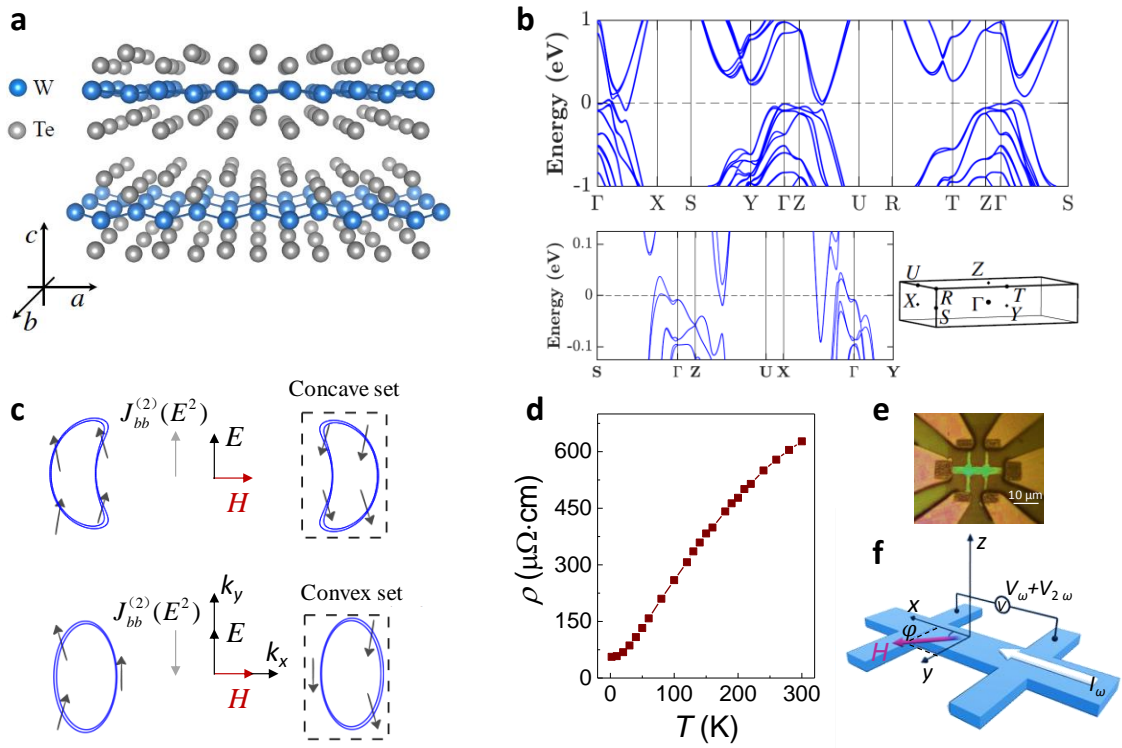

Figure 1

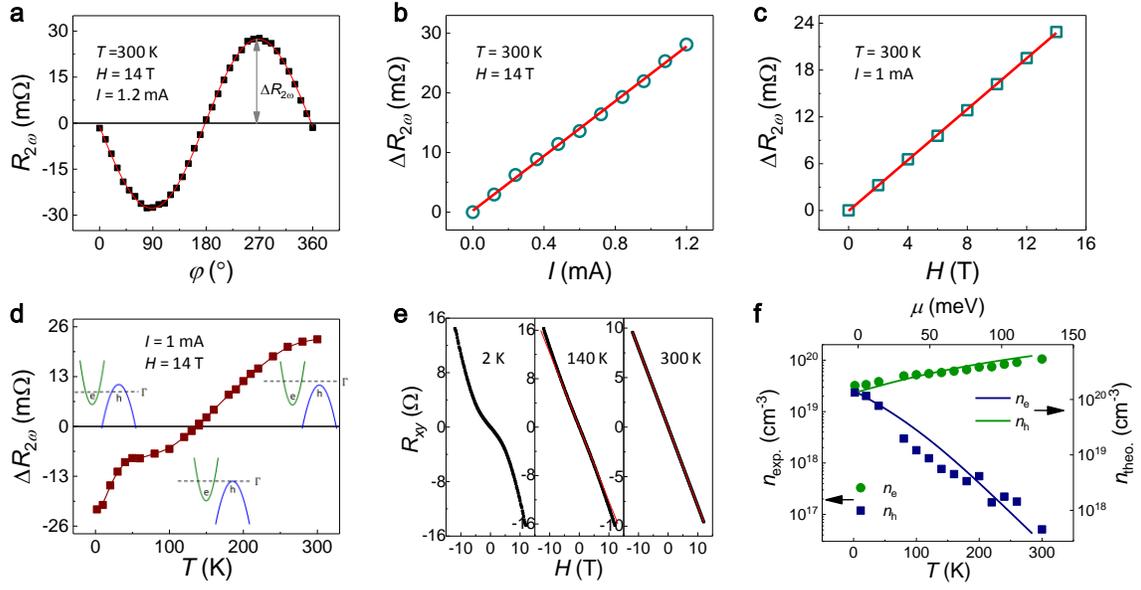

Figure 2

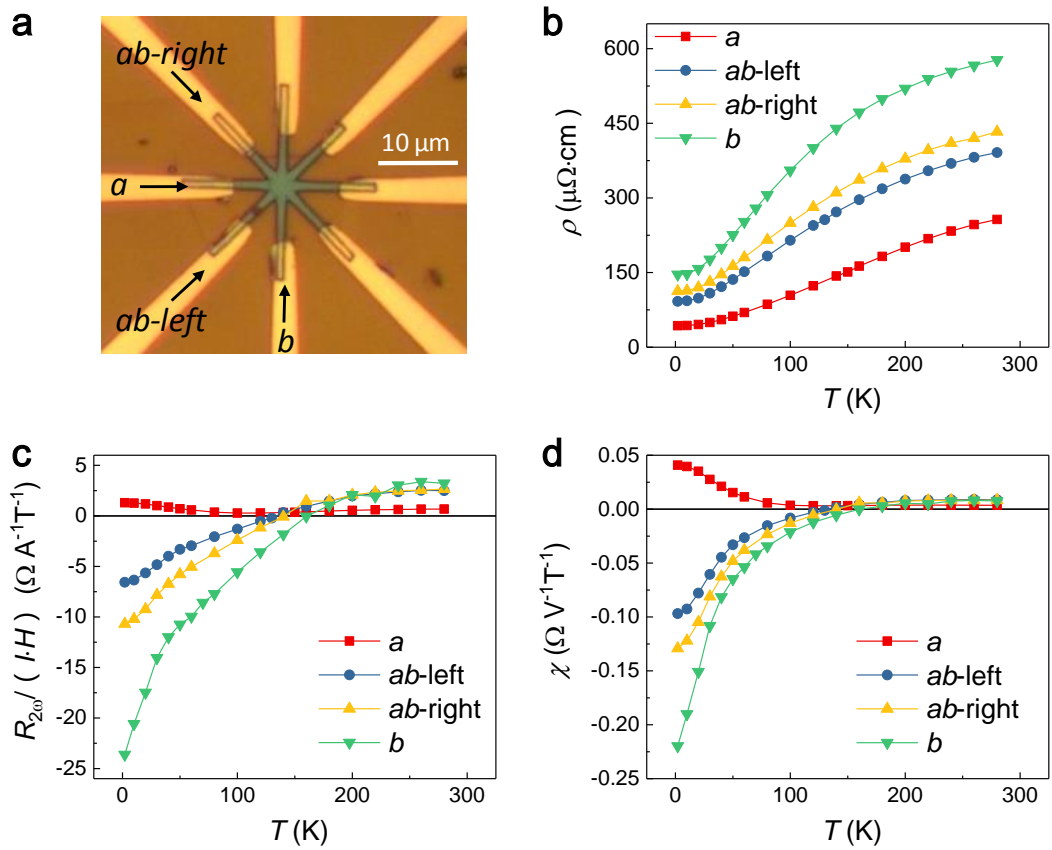

Figure 3

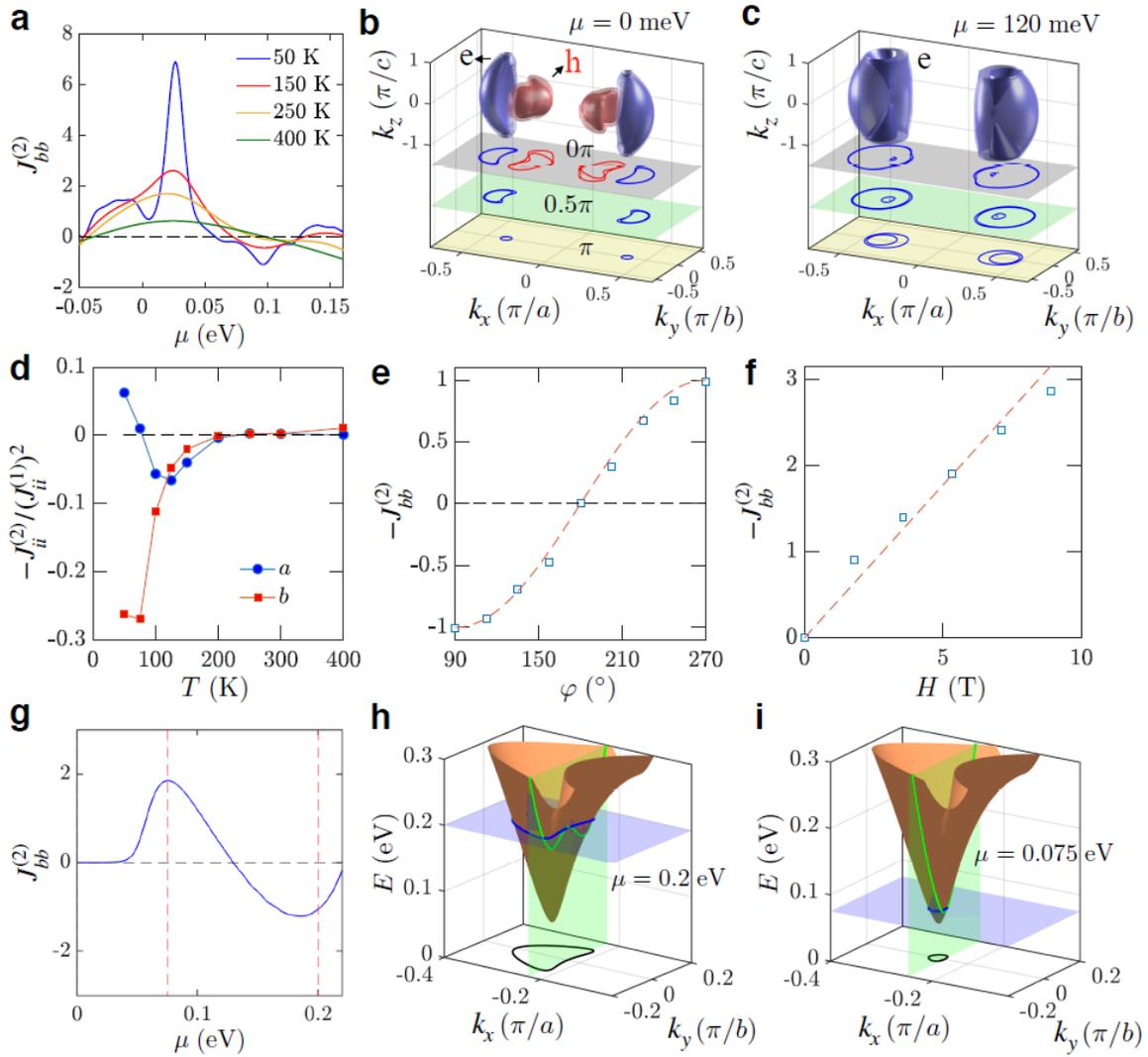

Figure 4